\documentclass[runningheads]{llncs}

\usepackage{graphicx}
\usepackage{hyperref}
\usepackage{amsmath}
\usepackage{amsthm}
\usepackage{amsfonts}
\usepackage{color}
\usepackage{multirow}
\usepackage{tabularx,booktabs}
\usepackage{xspace} 
\usepackage{caption}
\usepackage{subcaption}
\usepackage{enumitem}
\newcommand{\todo}[1]{}
\renewcommand{\todo}[1]{{\color{red}{#1}}}

\theoremstyle{definition}
\newtheorem{observation}[theorem]{Observation}

\begin{document}
\title{Seed-driven Document Ranking for Systematic Reviews: A Reproducibility Study}
\titlerunning{SDR for Systematic Reviews: A Reproducibility Study}
% If the paper title is too long for the running head, you can set
% an abbreviated paper title here
%
\author{Shuai Wang\orcidID{0000-0002-0726-5250} \and
Harrisen Scells\orcidID{0000-0001-9578-7157} \and\\
Ahmed Mourad\orcidID{0000-0002-9423-9404} \and
Guido Zuccon\orcidID{0000-0003-0271-5563}}
%
%\authorrunning{F. Author et al.}
% First names are abbreviated in the running head.
% If there are more than two authors, 'et al.' is used.
%
\institute{The University of Queensland, St. Lucia, Australia
	\email{\{shuai.wang2,h.scells,a.mourad,g.zuccon\}@uq.edu.au}}
%Springer Heidelberg, Tiergartenstr. 17, 69121 Heidelberg, Germany
%\email{lncs@springer.com}\\
%\url{http://www.springer.com/gp/computer-science/lncs} \and
%ABC Institute, Rupert-Karls-University Heidelberg, Heidelberg, Germany\\
%\email{\{abc,lncs\}@uni-heidelberg.de}}
%
\maketitle              % typeset the header of the contribution
\begin{abstract}

Screening or assessing studies is critical to the quality and outcomes of a systematic review. Typically, a Boolean query retrieves the set of studies to screen. As the set of studies retrieved is unordered, screening all retrieved studies is usually required for high-quality systematic reviews. 
%Screening all of the unordered retrieved studies limits the pace at which downstream activities of a systematic review can begin, as finding all relevant studies might occur at the end of the screening process. 
Screening prioritisation, or in other words, ranking the set of studies, enables downstream activities of a systematic review to begin in parallel.
We investigate a method that exploits seed studies -- potentially relevant studies used to seed the query formulation process -- for screening prioritisation. 
%Seed studies are potentially relevant studies used to seed the query formulation process. 
Our investigation aims to reproduce this method to determine if it is generalisable on recently published datasets and determine the impact of using multiple seed studies on effectiveness.
%, two aspects that we believe warrant a follow-up investigation from the original paper.
%
We show that while we could reproduce the original methods, we could not replicate their results exactly. However, we believe this is due to minor differences in document pre-processing, not deficiencies with the original methodology.
Our results also indicate that our reproduced screening prioritisation method, (1) is generalisable across datasets of similar and different topicality compared to the original implementation, (2) that when using multiple seed studies, the effectiveness of the method increases using our techniques to enable this, (3) and that the use of multiple seed studies produces more stable rankings compared to single seed studies. Finally, we make our implementation and results publicly available at the following URL: \url{https://github.com/ielab/sdr}.

\vspace{-12pt}
\end{abstract}

\section{Introduction}
\vspace{-12pt}
A systematic review is a focused literature review that synthesises all relevant literature for a specific research topic. Identifying relevant publications for medical systematic reviews is a highly tedious and costly exercise, often involving multiple reviewers to screen (i.e., assess) upwards of tens of thousands of studies. It is a standard practice to screen each study retrieved for a systematic review by a Boolean query. However, in recent years, there has been a dramatic rise in Information Retrieval methods that attempt to re-rank this set of studies for a variety of reasons, such as stopping the screening early (once a sufficient number of studies have been found) or beginning downstream phases of the systematic review process earlier (such as the acquisition of the full-text of studies). However, a known problem with many of these methods is that they use a different query from the Boolean query used to perform the initial literature search. Instead, most methods typically resort to less informationally representative sources for queries that can be used for ranking, such as the title of the systematic review, e.g.,~\cite{miwa2014reducing} (containing narrow information about the retrieval topic), or concatenating the clauses of the Boolean query together, e.g.,~\cite{alharbi2017ranking} (negating the structural information in Boolean clauses). We instead turn our attention to methods that use more informative sources of information to perform re-ranking. 

Indeed, we focus this reproducibility study on one such method: seed-driven document ranking (SDR) from Lee and Sun~\cite{lee2018seed}. SDR exploits studies that are known a priori to develop the research focus and search strategy for the systematic review. These studies are often referred to as `seed studies' and are commonplace in the initial phases of the systematic review creation process. This method and others such as CLF~\cite{scells2020clf} (which directly uses the Boolean query for ranking) have been shown to significantly outperform other methods that use a na\"ive query representation. Despite this, the SDR method was published when there was little data for those seeking to research this topic, and there have been methods published since that did not include SDR as a comparison. To this end, we devise the following research questions (RQs) to guide our investigation into why we are interested in reproducing the SDR method:

% limitations: one dataset, single seed study

%\subsection{Research Questions}
\vspace{-8pt}
\begin{description}[leftmargin=4pt]
\item[RQ1] \textit{Does the effectiveness of SDR generalise beyond the CLEF TAR 2017 dataset?} The original study was only able to be investigated on a single dataset of systematic review topics. In this study, we plan to use our replicated implementation of SDR to examine the effectiveness of this method across more recent datasets, and datasets that are more topically varied (CLEF TAR 2017 only contains systematic reviews about diagnostic test accuracy).
%While these are indeed a challenging kind of systematic review to produce and of great importance to develop methods to assist with their creation, there are many other kinds of systematic reviews of varying difficulty to produce).
\item[RQ2] \textit{What is the impact of using multiple seed studies collectively on the effectiveness of SDR?} The original study focused on two aspects of their method: an initial ranking using a single seed study and an iterative ranking which further uses the remaining seed studies one at a time. We focus on investigating the first aspect concerning the impact of multiple seed studies (multi-SDR) used collectively for input to produce an initial ranking.
%The original study only utilised a single seed study to produce a ranking.
\item[RQ3] \textit{To what extent do seed studies impact the ranking stability of single- and multi-SDR?} 
%Finally, we examine the impact of seed studies on the variability of SDR effectiveness. 
In a recent study by Scells et al.~\cite{scells2020comparison} to generate Boolean queries from seed studies, it was found that seed studies can have a considerable and significant effect on the effectiveness of resulting queries. We perform a similar study that aims to measure the variance in effectiveness of SDR in single- and multi- seed study settings.
\end{description}

%\noindent
%With the investigation into the above research questions, we will:
%\begin{itemize}[leftmargin=*]
%\item Demonstrate the \textbf{novelty} of the method by performing experiments on more datasets (\textbf{RQ1}), and experiments that reveal more about the effectiveness of the method (\textbf{RQ2}, \textbf{RQ3}).
%\item Assess the \textbf{impact} of SDR towards the Information Retrieval community and the wider systematic review community.
%\item Investigate the \textbf{reliability} of SDR by comparing it to several baselines, including those by participants at the CLEF TAR shared tasks, on publicly available datasets.
%\item Make our complete reproduced implementation of SDR publicly \textbf{available} for others to use as a baseline in future work on re-ranking for systematic reviews.
%\end{itemize}
\noindent
With the investigation into the above research questions, we will (1) demonstrate the \textbf{novelty} of the method by performing experiments on more datasets (\textbf{RQ1}), and experiments that reveal more about the effectiveness of the method (\textbf{RQ2}, \textbf{RQ3}), (2) assess the \textbf{impact} of SDR towards the Information Retrieval community and the wider systematic review community, (3) investigate the \textbf{reliability} of SDR by comparing it to several baselines on publicly available datasets, and (4) make our complete reproduced implementation of SDR publicly \textbf{available} for others to use as a baseline in future work on re-ranking for systematic reviews.

\vspace{-12pt}
\section{Replicating SDR}
\vspace{-8pt}

%\subsection{Observations on Relevant Documents}
In the original paper of Lee and Sun, they devise two experimental settings for SDR: an initial ranking of retrieved studies using a seed study and iteratively re-ranking by updating the query used for SDR with one seed study at a time to simulate the manual screening process. We focus on the initial ranking stage for two reasons: (1) screening prioritisation is an accepted practice in the systematic review creation process as all studies must still be screened~\cite{chandler2019cochrane}; and (2) an effective initial ranking will naturally result in a more effective and efficient re-ranking of studies, as more studies that are relevant will be identified faster.
The intuition for SDR is that relevant studies are similar to each other. The original paper makes two important observations about seed studies to support this intuition: (1) that relevant studies are more similar to each other than they are to non-relevant studies; and (2) that relevant studies share many \textit{clinical terms}. These two observations are used to inform the representation and scoring of studies, given a seed study. We attempt to replicate these observations below to verify both that our implementation follows the same steps to make similar observations and whether the assumptions derived from them hold.
%
%We begin by investigating these observations about clinical terms to determine if the intuitions about using them are valid or not.

\begin{observation}
For a given systematic review, its relevant documents share higher pair-wise similarity than that of irrelevant documents.
\end{observation}

\begin{figure}[t!]
	\vspace{-12pt}
	\centering
	\begin{minipage}[t]{.48\columnwidth}
		\includegraphics[width=\columnwidth]{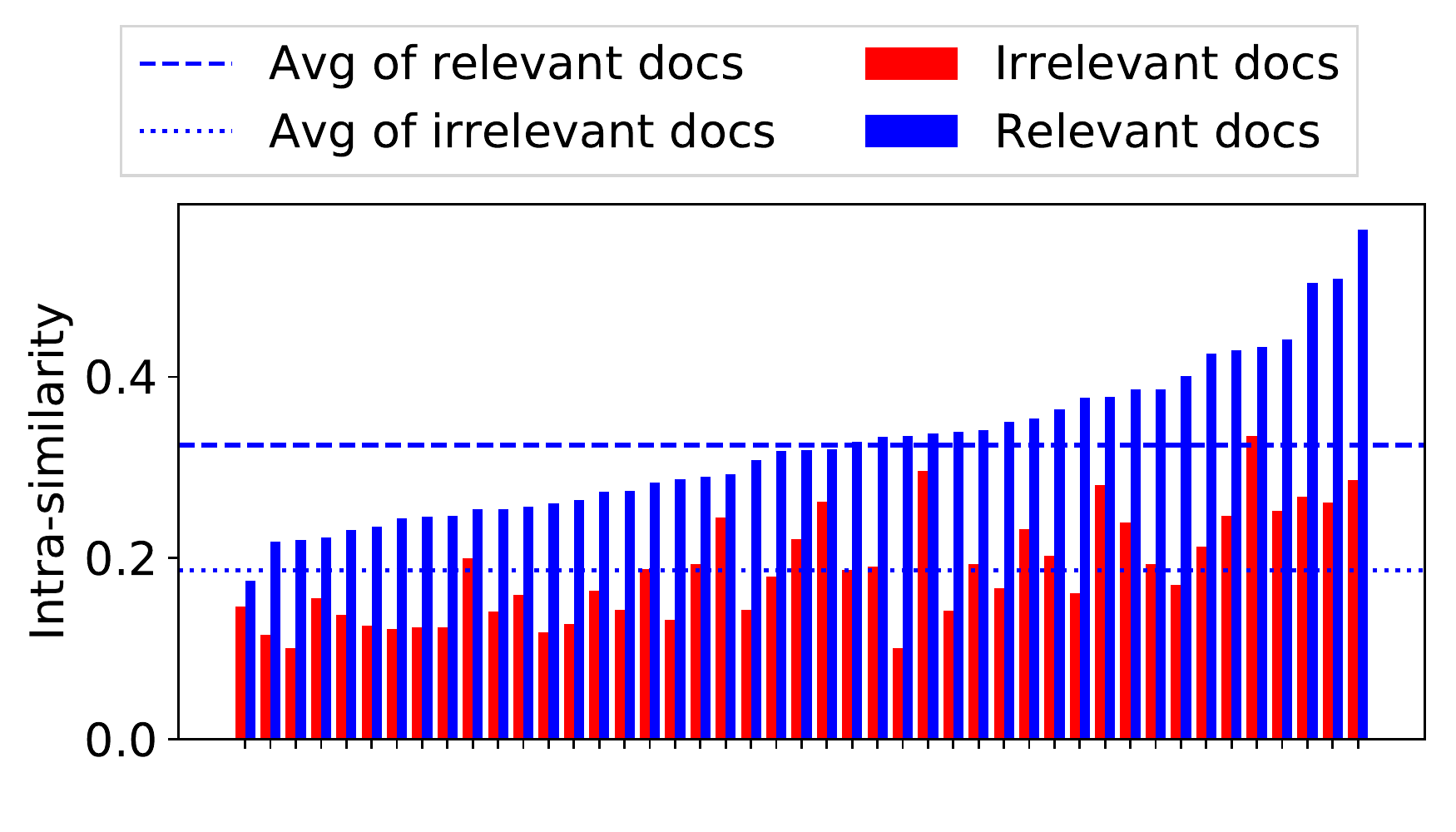}
		\vspace{-24pt}
		\caption{Intra-similarity between relevant studies and irrelevant studies.}
		\label{fig:replicating.pairwise-similarity}
	\end{minipage}
	\hspace{8pt}
	\begin{minipage}[t]{.48\columnwidth}
		\includegraphics[width=\columnwidth]{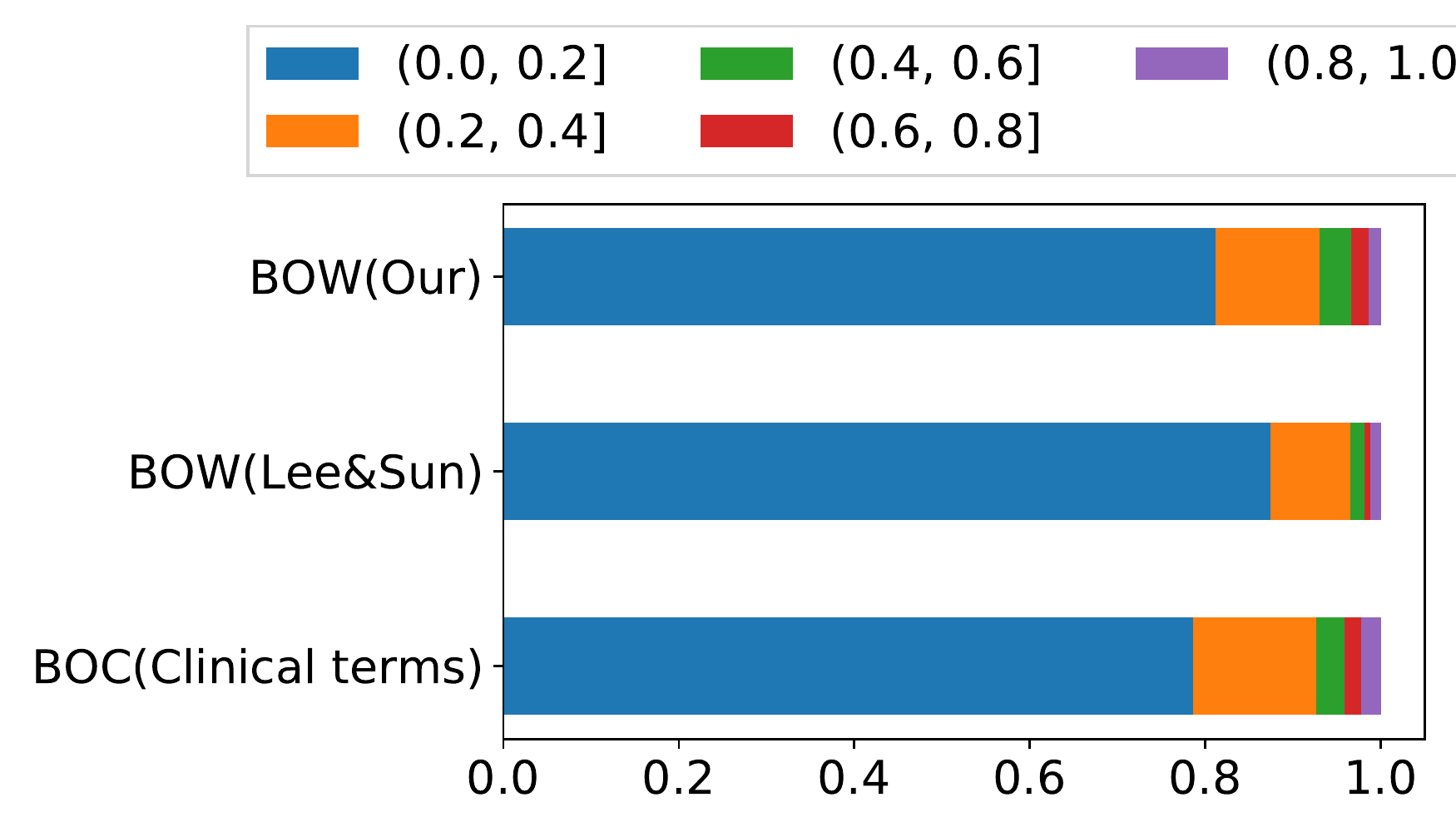}
		\vspace{-24pt}
		\caption{Distribution of terms in relevant studies.}
		\label{fig:replicating.commonality}
	\end{minipage}
	\vspace{-12pt}
\end{figure}

%The observation made is valid as relevant studies do share higher intra similarity compared with irrelevant documents. However, as negative studies were randomly chosen, the differences between their actual differences can only be treated as a reference, the actual comparing space for negative studies are much larger than that of positive studies, which means, higher intra-similarities with a particular seed document would always represent the study will be included definitely.
\noindent
We find that this observation is valid in our reproduction, as demonstrated by Figure~\ref{fig:replicating.pairwise-similarity}. In order to produce this plot, irrelevant studies were randomly under-sampled ten times. The number of non-relevant studies is always the same as the number of relevant studies for each topic. This means it is unlikely that we will produce the exact result initially found for this observation by Lee and Sun. Furthermore, one reason that the average pairwise similarity for the relevant studies may not match the original results is that the textual content of studies on PubMed may have changed or been updated. Rather than using a dump of PubMed from 2017, we used the latest version of studies on PubMed, as it is unknown the exact date that studies were extracted from PubMed in the original paper, and the CLEF TAR dataset does not give an exact date.

\begin{observation}
Relevant documents for a given systematic review share high commonality in terms of clinical terms.
\end{observation}

\noindent
We found that this observation is also valid in our reproduction, as demonstrated in Figure~\ref{fig:replicating.commonality}. It can be seen that the commonality of terms for the bag of words (BOW) and bag of clinical words (BOC) representations closely match those reported by Lee and Sun. However, we also found that with some minor modifications to the pre-processing of studies, we achieved a similar (yet still lower) commonality for terms using the BOW representation. We believe that the BOC representation shares a higher commonality of terms because the vocabulary size is smaller than the BOW representation. 
Naturally, with a smaller vocabulary, it is more likely for studies to share common terms.
%When pre-processing studies using the method suggested in original paper, we obtain a vocabulary size of 1,612,974 from BOW, thus the number of distinct clinical terms extracted only count for 4.6 \% of the vocabulary, while using preprocessing of proposed method in our paper, the number of distinct clinical terms counts for 30 \% in the BOW word space. With thee same number of BOC clinical terms extracted, our method only obtain 14.75 \% of the vocabulary size in original preprocessing method. 
When pre-processing studies using the method described in original paper, we find that BOC terms count for 4.6\% of the vocabulary, while they account for 31.2\% using our pre-processing. In fact, our BOW vocabulary is only 14.8\% their BOW vocabulary. Note that BOC is a distinct subset of BOW.

\vspace{-12pt}
\subsection{Document Representation}
\vspace{-4pt}

Given Observation 1 about relevant studies for this task, Lee and Sun chose to represent studies as a `bag of clinical words' (BOC). They chose to use the Unified Medical Language System (UMLS) as their ontology of clinical terms. UMLS is an umbrella ontology that combines many common medical ontologies such as SNOMED-CT and MeSH. In order to identify UMLS concepts (and therefore the clinical terms) within the studies, Lee and Sun combine the outputs of the NCBO Bioportal~\cite{noy2009bioportal} API\footnote{\url{http://data.bioontology.org/documentation}} and QuickUMLS~\cite{soldaini2016quickumls}. We follow their process as described, however we are not aware if it is not possible to set a specific version for the NCBO API. We use QuickUMLS version 1.4.0 with UMLS 2016AB.

\vspace{-12pt}
\subsection{Term Weighting}
\label{sec:replicate.term-weighting}
\vspace{-4pt}

SDR weights terms based on the intuition that terms in relevant studies are more similar to each other (or occur with each other more frequently) than non-relevant studies.
%Therefore, SDR weights terms to promote this intuition.
The weight of an individual term in a seed study is estimated by measuring to what extent it separates similar (pseudo-relevant) and dissimilar (pseudo-non-relevant) studies.
Formally, each term $t_i$ in a seed document $d_s$ ($t_i\in d_s$) is weighted using the function $\varphi (t_i,d_s) = \text{ln}\left(1+\frac{\gamma(D_{t_i},d_s)}{\gamma(D_{\bar{t_i}},d_s)}\right)$, 
%in Equation~\ref{eq:term-weight}
%\vspace{-8pt}
%\begin{equation}\label{eq:term-weight}
%	\phi (t_i,d_s) = \text{ln}\left(1+\frac{\gamma(D_{t_i},d_s)}{\gamma(D_{\bar{t_i}},d_s)}\right)
%\end{equation}
%
%\noindent
where $D_{t_i}$ represents the subset of candidate studies to be ranked where $t_i$ appears, and $D_{\bar{t_i}}$ represents the subset of candidate studies to be ranked where $t_i$ does not appear. The average similarity between studies is computed as $\gamma(D, d_s) = \frac{1}{|D|}\sum_{d_j\in D} sim(d_j, d_s)$, 
% in Equation~\ref{eq:similarity}:
%
%\vspace{-8pt}
%\begin{equation}\label{eq:similarity}
%	\gamma(D, d_s) = \frac{1}{|D|}\sum_{d_j\in D} sim(d_j, d_s)
%\end{equation}
%
%\noindent
where $sim$ is the cosine similarity between the vector representations of the candidate study $d_j$ and the seed study $d_i$. 
%In the original implementation, studies were represented as tf-idf vectors. For reproducibility purposes, we use the same representation.
We follow the original implementation and represent studies as tf-idf vectors.

\vspace{-12pt}
\subsection{Document Scoring}
\vspace{-4pt}

The original SDR implementation uses the query likelihood language model with Jelenik-Mercer smoothing for scoring studies. Typically, this ranking function is derived as indicated by QLM shown in Equation~\ref{eq:qlm-jm-sdr}, 
%
%\begin{equation}\label{eq:qlm-jm}
%	score(d,q) = \sum_{t_i\in d,q} c(t_i, q) \cdot \text{log}\left(1+\frac{1-\lambda}{\lambda}\cdot\frac{c(t_i,d)}{L_d\cdot p(t_i|\mathbb{C})}\right)
%\end{equation}
%
%\noindent
where $c(t_i, d_s)$ represents the count of a term in a seed study, $c(t_i, d)$ represents the count of a term in a candidate study, $L_d$ represents the number of terms in a study, $p(t_i|\mathbb{C})$ represents the probability of a term in a background collection, and $\lambda$ is the Jelenik-Mercer smoothing parameter.
To incorporate the term weights as described in Subsection~\ref{sec:replicate.term-weighting}, the original paper includes $\varphi$ function into the document scoring function as shown in Equation~\ref{eq:qlm-jm-sdr}:
\vspace{-8pt}
\begin{equation}\label{eq:qlm-jm-sdr}
	score(d,d_s) = \sum_{t_i\in d,d_s} 
	\overbrace{\varphi(t_i,d_s)}^\text{Term Weight}
	\cdot 
	\overbrace{c(t_i, d_s) \cdot \text{log}\left(1+\frac{1-\lambda}{\lambda}\cdot\frac{c(t_i,d)}{L_d\cdot p(t_i|\mathbb{C})}\right)}^\text{QLM}
\end{equation}

\noindent
where $p(t_i|\mathbb{C})$ is estimated using maximum likelihood estimation over the entire candidate set of studies $C$. In the original paper, when additional seed studies were ranked in the top-$k$ set of candidate seed studies (denoted as $d_{s^\prime}$), a re-ranking was initiated by expanding each $t_i$ in $d_s$ with the new terms from $d_{s^\prime}$. For our replication study, we only consider the initial ranking of candidate studies, as an abundance of baseline methods can be used as a comparison for this task. It is also arguably the most important step as a poor initial ranking will naturally result in a less effective and less efficient re-ranking.
\vspace{-12pt}
\subsection{Multi-SDR}
\vspace{-8pt}

One assumption in the original paper is that only a single seed study can be used at a time for ranking candidate studies. We propose a modification by studying the impact of using multiple seed studies collectively. In practice, it is common for Boolean queries (i.e., the search strategies used to retrieve the set of candidate studies we use for ranking) to be developed with a handful of seed studies, not just a single seed study. We hypothesise that the effectiveness of SDR will increase when multiple seed studies are used. 
Each relevant study must be used as a seed study for ranking, as the seed studies are not known in any of the collections we used. Therefore the average performance across topics was recorded (i.e., leave-one-out cross-validation). This study follows the methodology for the single-SDR method described in the subsections above. How we adapt single-SDR for a multi-SDR setting, and how we make this comparable to single-SDR is described as follows.
\vspace{-12pt}
\subsubsection{Grouping Seed Studies}
To study multi-SDR, we adopt a similar approach to the original paper; however, we instead randomly group multiple seed studies together and perform leave-one-out cross-validation over these groups. To account for any topic differences that may impact performance, we use a sliding window across the list of seed studies so that a seed study can appear in multiple groups. The number of seed studies to fill each group was chosen to be 20\% of the total seed studies. Rather than use a fixed number of seed studies, choosing different proportions simulates the use of seed studies in practice, i.e., different amounts of seed studies may be known before conducting a review. %We also ensure that topics where 20\% of the relevant studies results in a single seed study are not included.% so that our results only contain multi-SDR, and are not confounded by single-SDR results. \todo{the last sentence isn't clear to me}
%However, to account for differences in seed studies, we also shuffle the groups $k$ times and record the average performance across the $k$ shuffles. 
\vspace{-12pt}
\subsubsection{Combining Seed Studies for Multi-SDR}
The way we exploit multiple seed studies for SDR is, we believe, similar to how Lee and Sun used multiple seed studies in their relevance feedback approach to SDR. We concatenate seed studies together such that the resulting representation can be used directly with the existing single-SDR framework. We acknowledge that there may be more sophisticated approaches to exploit multi-SDR. However, we leave this as future work as it is out of the scope for this reproducibility study.

When computing term weights for multi-SDR, we also encountered computational infeasibility for large groups of seed studies. To this end, we randomly under-sampled the number of irrelevant studies to 50 each time we compute $\varphi$.
\vspace{-12pt}
\subsubsection{Comparing Single-SDR to Multi-SDR}
\label{sec:experimental-setup.comparing}

Directly comparing the results of multi-SDR to single-SDR is not possible due to the leave-one-out cross-validation style of evaluation used for single-SDR. To address this, we apply an oracle to identify the most effective single-SDR run out of all the seed studies used for a given multi-SDR run in terms of MAP. We then remove the other seed studies used in the multi-SDR run from the oracle-selected single-SDR run so that both runs share the same number of candidate studies for ranking. 
%Within each group of seed studies, we make an assumption that not all seed studies must be used for SDR. In practice, this may occur, for example, if a seed study is found to be indeed non-relevant to the topic of the systematic review when formulating the Boolean search. Therefore, we propose the following methods for selecting seed studies to be used for Multi-SDR:

%\begin{description}
%\item[Multi-All] A baseline method that does not exclude any seed studies. We use this method as a control to determine if using fewer seed studies can be more effective than using more seed studies.
%\item[Multi-Oracle] This method tries all combinations of the seed studies for SDR to identify the most effective combination. 
%\item[Multi-Diversity] This method first attempts to rank the seed studies in terms of diversity, and then takes the top-$k$ to be used in SDR, where $k$ is proportional to the number of seed studies. 
%\end{description}

\vspace{-12pt}
\section{Experimental Setup}
\vspace{-4pt}

\vspace{-8pt}
\subsection{Datasets}
\vspace{-4pt}

When the original SDR paper was published, only a single collection with results of baseline method implementations was available. 
We intend to assess the generalisability of their SDR method on several new collections which have been released since. The collections we consider are:
%In the five years since the SDR paper was published, several new collections have been made available that we intend to replicate SDR on to assess the methods' generalisability. The collections that we have chosen to replicate SDR on include:

\begin{description}[leftmargin=8pt]
	\item[CLEF TAR 2017~\cite{kanoulas2017clef}] This is the original dataset that was used to study SDR. We include this dataset to confirm that we achieve the same or similar results as the original paper. This collection includes 50 systematic review topics on diagnostic test accuracy -- a type of systematic review that is challenging to create. The 50 topics are split into 20 training topics and 30 testing topics. In our evaluation, we removed topics CD010653, CD010771, CD010386, CD012019, CD011549 as they contained only a single or no relevant studies to use as seed studies. For our experiments using multiple seed studies, we further removed topics CD010860, CD010775, CD010896, CD008643, CD011548, CD010438, CD010633, CD008686 due to low numbers of relevant studies.
	\item[CLEF TAR 2018~\cite{kanoulas2018clef}] This collection adds 30 diagnostic test accuracy systematic reviews as topics to the existing 2017 collection; however, it also removes eight because they are not `reliable for training or testing purposes. In total, this collection contains 72 topics. Our evaluation only used 30 additional reviews of the 2018 dataset and removed topics CD012216, CD009263, CD011515, CD011602, and CD010680 as they contained only a single or no relevant studies to use as seed studies. We also removed topic CD009263 because we ran into memory issues when running experiments on this topic due to many candidate documents (approx. 80,000). For our experiments using multiple seed studies, we removed topics CD012083, CD012009, CD010864, CD011686, CD011420 due to low numbers of relevant studies.
	\item[CLEF TAR 2019~\cite{kanoulas2019clef}] This collection further develops on the previous years' by also including systematic reviews of different types. From this collection, we use the 38 systematic reviews of interventions (i.e., a different type of diagnostic test accuracy).\footnote{Although the overview paper claims there are 40 interventions topics, there are two topics that appear in both training and testing splits. However, like the previous datasets, we ignore these splits and combine the training and testing splits.} We use this collection to study the generalisability of SDR on other kinds of systematic reviews. In our evaluation, we removed topics CD010019, CD012342, CD011140, CD012120, CD012521 as they contained only a single or no relevant studies to use as seed studies. For our experiments using multiple seed studies, we further removed topics CD011380, CD012521, CD009069, CD012164, CD007868, CD005253, CD012455 due to low numbers of relevant studies.
	%\item[SR'17~\cite{scells2017collection}] This collection contains 125 topics of intervention systematic reviews. We also use this collection to investigate the generalisability of SDR on systematic reviews other than diagnostic test accuracy.
\end{description}

\vspace{-22pt}
\subsection{Baselines}
\vspace{-4pt}

The baselines in the original paper included the best performing method from the CLEF TAR 2017 participants, several seed-study-based methods, and variations of the scoring function used by SDR.
For our experiments, we compare our reproduction of SDR to all of the original baselines that we have also reproduced from the original paper.%, in addition to the results of others on each of the datasets, where available (i.e., participants to the CLEF TAR shared tasks).
The baselines in the original paper include: BM25-\{BOW,BOC\}, QLM-\{BOW,BOC\}, SDR-\{BOW,BOC\}, and AES-\{BOW,BOC\}. The last method, AES, is an embedding-based method that averages the embeddings for all terms in the seed studies. The AES method uses pre-trained word2vec embeddings using PubMed and Wikipedia (as specified in the original paper). We also include a variation that uses only PubMed embeddings (AES-P). Finally, we also include the linear interpolation between SDR and AES, using the same parameter as the original paper ($\alpha=0.3$). We use the same versions of the pre-trained embeddings as the original paper.

\vspace{-12pt}
\subsection{Evaluation Measures}
\vspace{-4pt}

For comparison to the original paper, we use the same evaluation measures. These include MAP, precision@k, recall@k, LastRel\%, and Work Saved over Sampling (WSS). LastRel is a measure introduced at CLEF TAR'17~\cite{kanoulas2017clef}. It is calculated as the rank position of the last relevant document. LastRel\% is the normalised percentage of studies that must be screened in order to obtain all relevant studies. Work Saved Over Sampling; a measure initially proposed to measure classification effectiveness~\cite{cohen2006reducing}, is calculated instead here, by computing the fraction of studies that can be removed from screening to obtain all relevant documents; i.e., $\text{WSS} = \frac{|C|-\text{LastRel}}{|C|}$.
Where $C$ is the number of studies originally retrieved (i.e., the candidate set for re-ranking). 
For precision@k and recall@k, we report much deeper levels of $k$: the original paper reported $k=\{10,20,30\}$; where we report $k=\{10,100,1000\}$. Furthermore, we also report nDCG at these k-values, as it provides additional information about relevant study rank positions.
We compute LastRel\% and WSS using the scripts used in CLEF TAR 2017. For all other evaluation measures we use \texttt{trec\_eval} (version 9.0.7).

\vspace{-12pt}
\subsection{Document Pre-Processing}
\vspace{-4pt}
It is widely known that document pre-processing (e.g., tokenisation, stopwords, or stemming) can have a profound effect on ranking performance~\cite{croft2002combining}. Although the original paper provides information about the versions of the libraries it uses for ranking, there were fewer details, such as how documents were tokenised or which stopword list was used. We reached out to the original authors to confirm the exact experimental settings. From the original paper, documents were split using space, then stopwords were removed using nltk.

The modifications we made to the document pre-processing pipeline were that documents were first pre-processed to remove punctuation marks and then tokenised using gensim version 3.2.0 tokeniser. For stopwords, as the original authors have not specified the nltk version, we used the latest version at the time of publishing, version 3.6.3. Then terms used are lowercased for in all methods except for AES. No stemming has been applied in either pre-processing pipeline.

\vspace{-8pt}
\vspace{-8pt}
\section{Results}
\vspace{-10pt}

Before we investigate the three research questions of our reproducibility study, we first examine the extent to which we were able to replicate the results of Lee and Sun. In this study, we were unable to exactly replicate the results due to what we believe to be minor differences in document pre-processing and evaluation setup. Despite these difference, the results in Table~\ref{table:results.single.2017} show a similar performance across the baselines and evaluation measures compared to what Lee and Sun originally reported in their paper for our pre-processing pipeline. %When we compare the performance of their pre-processing pipeline as described in their paper, we were unable to reproduce the same trend in effectiveness as in their paper.

The results observed comparing the document pre-processing pipeline for the BOW representation as described by Lee and Sun (*-LEE) to our document pre-processing pipeline show that the BOW baselines may not have been as strong as if the original authors had performed a similar pipeline as us. We find that although the results comparing their baseline is statistically significant with our best performing method, our baseline is not significantly different. Finally, we find that the SDR-BOW-AES-LEE method, which corresponds to their most effective method, is significantly worse than our most effective method for 2017, SDR-BOW-AES-P.
%There was also a significant impact on effectiveness when combining their SDR pre-processing with AES on the 2017 dataset.

In terms of the BOC representation we were unable to identify a more effective pipeline for extracting clinical terms. Here, we applied the clinical term extraction tools over \textit{individual terms} in the document (following the pre-processing of Lee and Sun), and not \textit{the entire document}. Although we find this to be counter-intuitive, as tools like QuickUMLS and the NCBO API use text semantics to match n-grams, the result of applying the tools to individual terms has the effect of reducing the vocabulary of a seed study to the key concepts.

Finally, comparing our evaluation setup to Lee and Sun, we find that there were a number of topics in the CLEF TAR 2017 dataset that were incompatible with SDR. Rather than attempting to replicate their results, we simply do not compare their original results with ours, since we do not have access to their run files or precise evaluation setup. Furthermore, when we compare the results we report from to the best performing participant at CLEF TAR 2017 that did not use relevance feedback~\cite{alharbi2017ranking}, we remove the same topics from the run file of this participant for fairness. Although this method cannot be directly compared to, we can see that even relatively unsophisticated methods that use seed studies such as BM25-BOW are able to outperform the method by this participant.

\vspace{-12pt}
\subsection{Generalisability of SDR}
\vspace{-10pt}

We next investigate the first research question: \textit{Does the effectiveness of SDR generalise beyond the CLEF TAR 2017 dataset?} In Table~\ref{table:results.generalisability}, we can see that the term weighting of SDR almost always increases effectiveness compared to using only QLM, and that interpolation with AES can have further benefits to effectiveness. However, we note that few of these results are statistically significant. 

While we are unable to include all of the results for space reasons, we find that SDR-BOC-AES-P was not always the most effective SDR method. Indeed on the 2019 dataset, SDR-BOW was the most effective.
%When comparing the results of the 2019 dataset to the 2017 and 2018 datasets that contain systematic reviews about diagnostic test accuracy, both share a similar trend in the effectiveness of methods.
%Furthermore, we note that the methods that combine SDR with AES generally achieve the highest performance across all methods, except for SDR-BOW on the 2019 dataset (however, these results are not statistically significant).
The reason for this may be due to the difference in topicality of the 2019 dataset. 
This suggests that not only is the method of identifying clinical terms not suitable for these intervention systematic review topics, but that the interpolation between SDR and AES may require dataset-specific tuning.

\begin{table}[t]
	\centering
	\tiny
	
\begin{tabular}{l|p{21pt}|p{19pt}p{19pt}p{21pt}|p{19pt}p{19pt}p{21pt}|p{19pt}p{19pt}p{21pt}|p{21pt}|p{19pt}}
	
	\toprule
	Method&MAP&Prec.&Prec.&Prec.&Recall&Recall&Recall&nDCG&nDCG&nDCG&LR\%&WSS\\
	&&10&100&1000&10&100&1000&10&100&1000&&\\\hline
    \addlinespace[2pt]
Sheffield-run-2\cite{alharbi2017ranking}&0.1706&0.1367&0.0703&0.0156&0.1759&0.5133&0.8353&0.2089&0.3342&0.4465&0.4660&0.5340\\\midrule
BM25-BOW-LEE&0.1710$^\dagger$&0.2027$^\dagger$&0.0867$^\dagger$&0.0195$^\dagger$&0.1543&0.5118$^\dagger$&0.8798$^\dagger$&0.2439$^\dagger$&0.3419$^\dagger$&0.4770$^\dagger$&0.4902$^\dagger$&0.5098$^\dagger$\\

BM25-BOW&0.1810&0.2128$^\dagger$&0.0898$^\dagger$&0.0200&0.1646&0.5232$^\dagger$&0.8928&0.2560&0.3534$^\dagger$&0.4899$^\dagger$&0.4427$^\dagger$&0.5573$^\dagger$\\
BM25-BOC&0.1764$^\dagger$&0.2145$^\dagger$&0.0895$^\dagger$&0.0200&0.1562&0.5215$^\dagger$&0.8944&0.2539&0.3496$^\dagger$&0.4871$^\dagger$&0.4401$^\dagger$&0.5599$^\dagger$\\\midrule

QLM-BOW-LEE&0.1539$^\dagger$&0.1846$^\dagger$&0.0778$^\dagger$&0.0184$^\dagger$&0.1367$^\dagger$&0.4664$^\dagger$&0.8508$^\dagger$&0.2198$^\dagger$&0.3091$^\dagger$&0.4454$^\dagger$&0.4662$^\dagger$&0.5338$^\dagger$\\
QLM-BOW&0.1973&0.2360&0.0964&0.0203&\textbf{0.1855}&0.5464&0.9081&\textbf{0.2827}&0.3772&0.5100&0.3851&0.6149\\
QLM-BOC&0.1894&0.2330&0.0951&0.0202&0.1809&0.5376&0.9032&0.2771&0.3684&0.5018&0.3936&0.6064\\\midrule
SDR-BOW-LEE&0.1533$^\dagger$&0.1777$^\dagger$&0.0780$^\dagger$&0.0185$^\dagger$&0.1304$^\dagger$&0.4710$^\dagger$&0.8576$^\dagger$&0.2142$^\dagger$&0.3088$^\dagger$&0.4460$^\dagger$&0.4660$^\dagger$&0.5340$^\dagger$\\
SDR-BOW&0.1972&0.2264&0.0952&0.0204&0.1718&0.5398&0.9083&0.2739&0.3728&0.5081&0.3742&0.6258\\
SDR-BOC&0.1953&0.2329&\textbf{0.0974}&\textbf{0.0206}&0.1751&0.5530&0.9151&0.2756&0.3751&0.5086&0.3689&0.6311\\\midrule

AES-BOW&0.1516$^\dagger$&0.1768$^\dagger$&0.0785$^\dagger$&0.0190$^\dagger$&0.1369$^\dagger$&0.4611$^\dagger$&0.8794$^\dagger$&0.2163$^\dagger$&0.3106$^\dagger$&0.4552$^\dagger$&0.4549$^\dagger$&0.5451$^\dagger$\\
AES-BOW-P&0.1604$^\dagger$&0.1872$^\dagger$&0.0809$^\dagger$&0.0193$^\dagger$&0.1480$^\dagger$&0.4954$^\dagger$&0.8895$^\dagger$&0.2274$^\dagger$&0.3255$^\dagger$&0.4669$^\dagger$&0.4088$^\dagger$&0.5912$^\dagger$\\\midrule

SDR-BOW-LEE-AES&0.1716$^\dagger$&0.2008$^\dagger$&0.0870$^\dagger$&0.0197&0.1484$^\dagger$&0.5250$^\dagger$&0.8988$^\dagger$&0.2389$^\dagger$&0.3429$^\dagger$&0.4792$^\dagger$&0.4148$^\dagger$&0.5852$^\dagger$\\
SDR-BOW-AES&0.1958&0.2309&0.0957&0.0203&0.1750&0.5568&0.9163&0.2756&0.3764&0.5090&0.3880$^\dagger$&0.6120$^\dagger$\\
SDR-BOC-AES&0.1964&0.2364&0.0972&0.0204&0.1770&0.5699&0.9195&0.2800&0.3813&0.5117&0.3830$^\dagger$&0.6170$^\dagger$\\\midrule

SDR-BOW-LEE-AES-P&0.1764$^\dagger$&0.2058$^\dagger$&0.0883$^\dagger$&0.0199&0.1570&0.5349$^\dagger$&0.9081$^\dagger$&0.2448$^\dagger$&0.3500$^\dagger$&0.4865$^\dagger$&0.3796$^\dagger$&0.6204$^\dagger$\\
SDR-BOW-AES-P&0.1983&0.2322&0.0961&0.0204&0.1740&0.5673&0.9206&0.2768&0.3812&0.5128&0.3608&0.6392\\
SDR-BOC-AES-P&\textbf{0.1984}&\textbf{0.2369}&0.0970&0.0205&0.1788&\textbf{0.5737}&\textbf{0.9241}&0.2807&\textbf{0.3837}&\textbf{0.5147}&\textbf{0.3566}&\textbf{0.6434}\\

\bottomrule
\end{tabular}

	\caption{Reproduction results of baselines and SDR methods on the CLEF TAR 2017 dataset. For BOW methods, the pre-processing pipeline used by Lee and Sun is denoted by `-LEE'. BOW methods that do not have this demarcation correspond to our pipeline. For AES methods, word2vec PubMed embeddings are denoted by `-P'. AES methods that do not have this demarcation correspond to word2vec embeddings that include PubMed and Wikipedia. Statistical significance (Student's two-tailed paired t-test with Bonferonni correction, $p<0.05$) between the most effective method (SDR-BOC-AES-P) and all other methods is indicated by $\dagger$.}
	\label{table:results.single.2017}
	\vspace{-28pt}
\end{table}

\begin{table}[t]
	\centering
	\tiny
	\begin{tabular}{ll|l|lll|lll|lll|l|l}
	
	\toprule
 & Method &MAP&Prec.&Prec.&Prec.&Recall&Recall&Recall&nDCG&nDCG&nDCG&LR\%&WSS\\
&&10&100&1000&10&100&1000&10&100&1000&&\\
\hline \addlinespace[2pt]

\multirow{3}{*}{\rotatebox{90}{2017}}&QLM&0.1894&0.2330&0.0951&0.0202&\textbf{0.1809}&0.5376&0.9032&0.2771&0.3684&0.5018&0.3936&0.6064\\
&SDR&0.1953&0.2329&\textbf{0.0974}&\textbf{0.0206}&0.1751&0.5530&0.9151&0.2756&0.3751&0.5086&0.3689&0.6311\\
& SDR-AES-P&\textbf{0.1984}&\textbf{0.2369}&0.0970&0.0205&0.1788&\textbf{0.5737}&\textbf{0.9241}&\textbf{0.2807}&\textbf{0.3837}&\textbf{0.5147}&\textbf{0.3566}&\textbf{0.6434}\\\midrule

\multirow{3}{*}{\rotatebox{90}{2018}}&QLM-BOC&0.2344&0.2594&0.1130&0.0219&0.1821&\textbf{0.6214}&0.9104&0.3141&0.4156&0.5312&0.3317$^\dagger$&0.6683$^\dagger$\\
&SDR&0.2374&0.2549&0.1136&0.0221&0.1798&0.6176&0.9174&0.3117&0.4163&0.5351&0.3024&0.6976\\
&SDR-AES-P&\textbf{0.2503}&\textbf{0.2688}&\textbf{0.1161}&\textbf{0.0222}&\textbf{0.1957}&0.6036&\textbf{0.9234}&\textbf{0.3259}&\textbf{0.4243}&\textbf{0.5445}&\textbf{0.2695}&\textbf{0.7305}\\\midrule

\multirow{3}{*}{\rotatebox{90}{2019}}& QLM&0.2614&0.2599&0.0881&0.0169&0.2748&0.7032&0.9297&0.3458&0.4700&0.5482&0.4085&0.5915\\
&SDR&0.2790&0.2663&\textbf{0.0899}&\textbf{0.0169}&\textbf{0.3048}&0.7151&0.9337&0.3594&0.4846&0.5602&0.3819&\textbf{0.6181}\\
&SDR-AES-P&\textbf{0.2827}&\textbf{0.2667}&0.0898&0.0168&0.2973&\textbf{0.7174}&\textbf{0.9378}&\textbf{0.3649}&\textbf{0.4913}&\textbf{0.5672}&\textbf{0.3876}&0.6124\\

	\bottomrule
\end{tabular}

	\caption{Generalisability of results on the CLEF TAR 2017, 2018 and 2019 datasets. Representations used in this table are all BOC.  Statistical significance (Student's two-tailed paired t-test with Bonferonni correction, $p<0.05$) between the most effective method (SDR-AES-P) and other methods is indicated by $\dagger$.}
	\label{table:results.generalisability}
%	\vspace{-12pt}
\end{table}

%\begin{table}[t]
%	\centering
%	\tiny
%	\input{results-table-2019.tex}
%	\caption{Reproduction results of baselines and SDR methods on the CLEF TAR 2019 dataset. For demarcations used in ranking model names and statistical significance details, refer to the caption in Table~\ref{table:results.single.2017}.}
%	\label{table:results.single.2019}
%	\vspace{-24pt}
%\end{table}

\vspace{-14pt}
\subsection{Effect of Multiple Seed Studies}

\vspace{-6pt}

\begin{table}[t]
	\centering
	\tiny
	
\begin{tabular}{ll|l|p{21pt}p{21pt}p{21pt}|p{21pt}p{21pt}p{21pt}|lll|l|l}
	
	\toprule
	 & Method &MAP&Prec.&Prec.&Prec.&Recall&Recall&Recall&nDCG&nDCG&nDCG&LR\%&WSS\\
	 & & &10&100&1000&10&100&1000&10&100&1000& & \\\hline \addlinespace[2pt]
\multirow{5}{*}{\rotatebox{90}{2017}} & Single-BOC &0.3116&0.4235&0.1463&0.0255&0.2219&0.6344&0.9469&0.4830&0.5330&0.6595&0.3699&0.6301\\
&Single-BOW&0.3098&0.4076&0.1465&0.0255&0.2158&0.6366&0.9472&0.4679&0.5312&0.6566&0.3687&0.6313\\
&Multi-BOC&0.4554$^\dagger$&0.5804$^\dagger$&0.1752$^\dagger$&0.0272$^\dagger$&0.2917$^\dagger$&0.7151$^\dagger$&0.9661$^\dagger$&0.6817$^\dagger$&0.6765$^\dagger$&0.7835$^\dagger$&0.3427&0.6573\\
&Multi-BOW&0.4610$^\dagger$&0.5910$^\dagger$&0.1762$^\dagger$&0.0272$^\dagger$&0.2951$^\dagger$&0.7155$^\dagger$&0.9659$^\dagger$&0.6924$^\dagger$&0.6805$^\dagger$&0.7866$^\dagger$&0.3450&0.6550\\\cmidrule{2-14}
&\% Change&47.4801&41.0234&20.0132&6.6705&34.1131&12.5557&2.0029&44.5398&27.5202&19.3035&-6.8792&4.0283\\\midrule
\multirow{5}{*}{\rotatebox{90}{2018}}&Single-BOC&0.3345&0.4443&0.1671&0.0285&0.2041&0.6181&0.9280&0.5011&0.5296&0.6551&0.2641&0.7359\\
&Single-BOW&0.3384&0.4433&0.1678&0.0286&0.2062&0.6197&0.9383&0.4955&0.5301&0.6579&0.2577&0.7423\\
&Multi-BOC&0.4779$^\dagger$&0.6130$^\dagger$&0.1979$^\dagger$&0.0307$^\dagger$&0.2821$^\dagger$&0.6997$^\dagger$&0.9592$^\dagger$&0.7199$^\dagger$&0.6823$^\dagger$&0.7908$^\dagger$&0.2394$^\dagger$&0.7606$^\dagger$\\
&Multi-BOW&0.4809$^\dagger$&0.6109$^\dagger$&0.1978$^\dagger$&0.0306$^\dagger$&0.2813$^\dagger$&0.6968$^\dagger$&0.9585$^\dagger$&0.7218$^\dagger$&0.6835$^\dagger$&0.7924$^\dagger$&0.2396&0.7604\\\cmidrule{2-14}
&\% Change&42.5011&37.8814&18.1509&7.2657&37.3377&12.8217&2.7561&44.6754&28.8870&20.5797&-8.1919&2.8990\\\midrule
\multirow{5}{*}{\rotatebox{90}{2019}}&Single-BOC&0.3900&0.4249&0.1285&0.0221&0.3196&0.7261&0.9368&0.5365&0.6164&0.6897&0.4304&0.5696\\
&Single-BOW&0.3925&0.4418&0.1272&0.0222&0.3366&0.7243&0.9386&0.5516&0.6164&0.6916&0.4285&0.5715\\
&Multi-BOC&0.5341$^\dagger$&0.5746$^\dagger$&0.1533$^\dagger$&0.0243$^\dagger$&0.3962$^\dagger$&0.7896$^\dagger$&0.9622$^\dagger$&0.7105$^\dagger$&0.7458$^\dagger$&0.8091$^\dagger$&0.3852$^\dagger$&0.6148$^\dagger$\\
&Multi-BOW&0.5374$^\dagger$&0.5864$^\dagger$&0.1521$^\dagger$&0.0244$^\dagger$&0.4031$^\dagger$&0.7853$^\dagger$&0.9616$^\dagger$&0.7223$^\dagger$&0.7466$^\dagger$&0.8114$^\dagger$&0.3877$^\dagger$&0.6123$^\dagger$\\\cmidrule{2-14}
&\% Change&36.9305&33.9958&19.3948&9.9327&21.8599&8.5825&2.5819&31.6927&21.0510&17.3213&-10.0189&7.5424\\

\bottomrule
\end{tabular}
	\caption{Results comparing single-SDR and multi-SDR on the CLEF TAR 2017, 2018, and 2019 datasets. Note that the results for single-SDR are not directly comparable to the above tables as explained in Section~\ref{sec:experimental-setup.comparing}. Statistical differences (Student's paired two-tailed t-test, $p<0.05$) are indicated pairwise between the single- and multi- SDR BOC and BOW methods for each year (e.g., single-SDR-BOC-AES-P vs. multi-SDR-BOC-AES-P for 2017). \% Change indicates the average difference between single- and multi-\{BOW+BOC\}.}
	\label{table:results.multi}
	\vspace{-16pt}
\end{table}

Next, we investigate the second research question: \textit{What is the impact of using multiple seed studies collectively on the effectiveness of SDR?} Firstly, several topics were further removed for these experiments. Therefore, the results of single-SDR in Table~\ref{table:results.multi} are not directly comparable to the results in Tables~\ref{table:results.single.2017} and~\ref{table:results.generalisability}. In order to measure the effect multiple studies has on SDR compared to single seed studies, we also remove the same topics for single-SDR.

We find that across all three datasets, compared to single-SDR, multi-SDR can significantly increase the effectiveness. We also find that the largest increases in effectiveness are seen on shallow metrics across all three CLEF TAR datasets. This has implications for the use of SDR in practice, as typically, multiple seed studies are available before conducting the screening process. Therefore, when multiple seed studies are used for the initial ranking process, active learning methods that iteratively rank unjudged studies will naturally be more effective (as more relevant studies are retrieved in the early rankings). However, we argue that the assumption that relevant studies are a good surrogate for seed studies made by Lee and Sun~\cite{lee2018seed} and by others in other work such as Scells et al.~\cite{scells2020comparison} may be weak and that methods that utilise relevant studies for this purpose overestimate effectiveness. In reality, seed studies may not be relevant studies. They may be discarded once a Boolean query has been formulated (e.g., they may not be randomised controlled trials or unsuitable for inclusion in the review).

\vspace{-12pt}
\subsection{Variability of Seed Studies on Effectiveness}
\vspace{-4pt}

\begin{figure}[t!]
\vspace{-12pt}
\begin{subfigure}[t!]{.329\columnwidth}
	\includegraphics[width=\columnwidth]{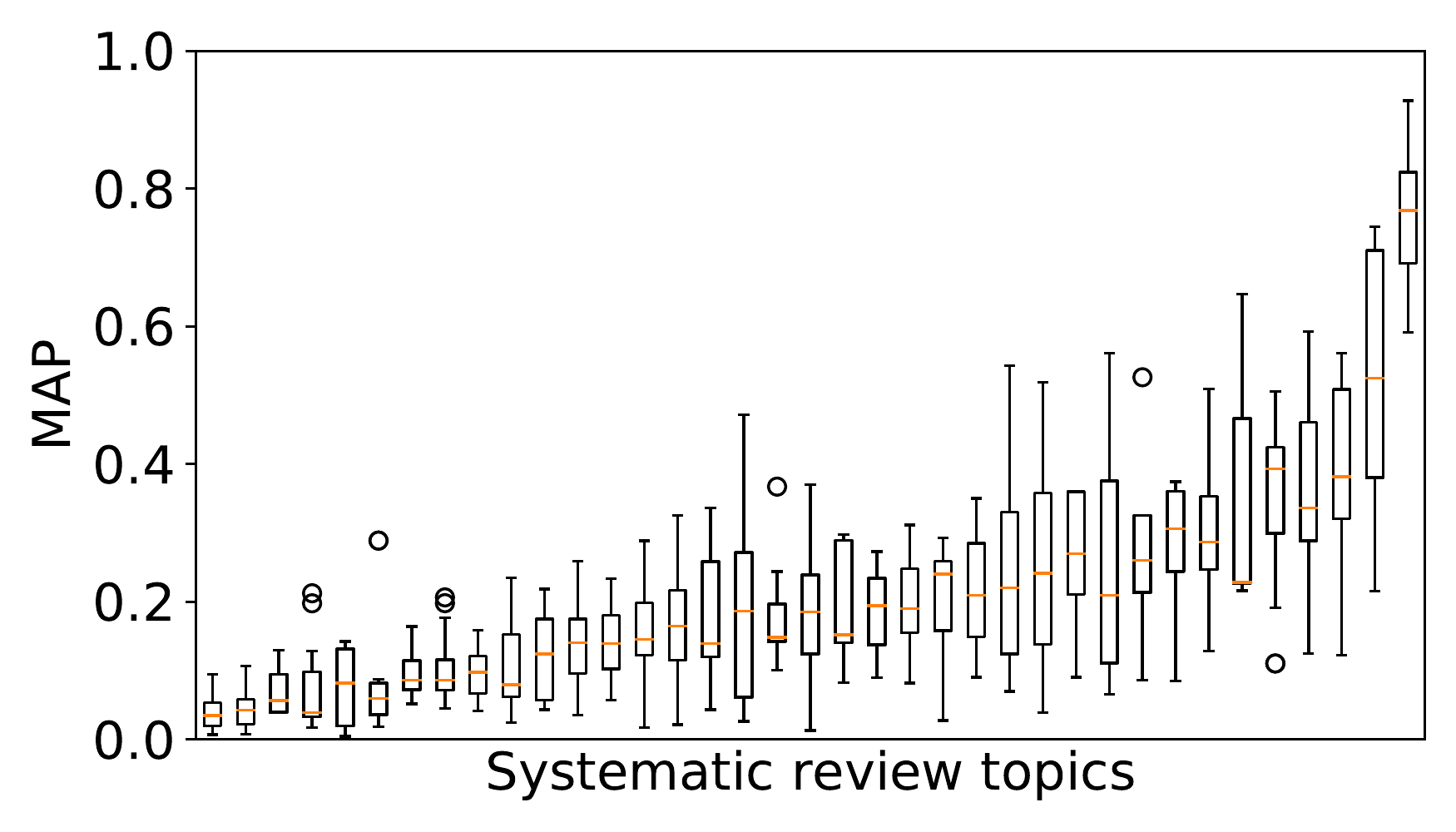}
	\caption{Single-SDR; 2017}
	\label{fig:results.2017var-single}
\end{subfigure}
\begin{subfigure}[t!]{.329\columnwidth}
	\includegraphics[width=\columnwidth]{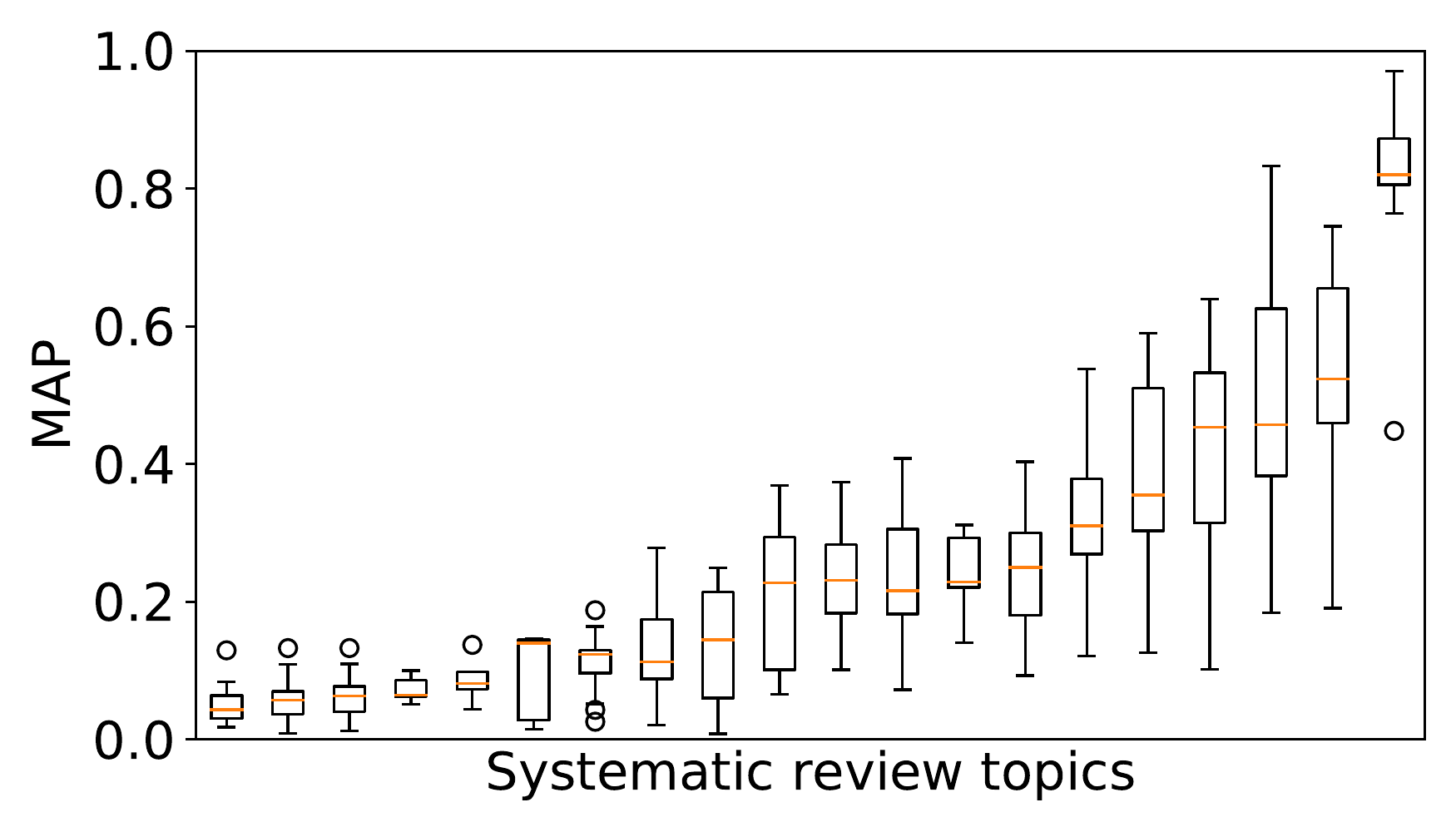}
	\caption{Single-SDR; 2018}
	\label{fig:results.2018var-single}
\end{subfigure}
\begin{subfigure}[t!]{.329\columnwidth}
	\includegraphics[width=\columnwidth]{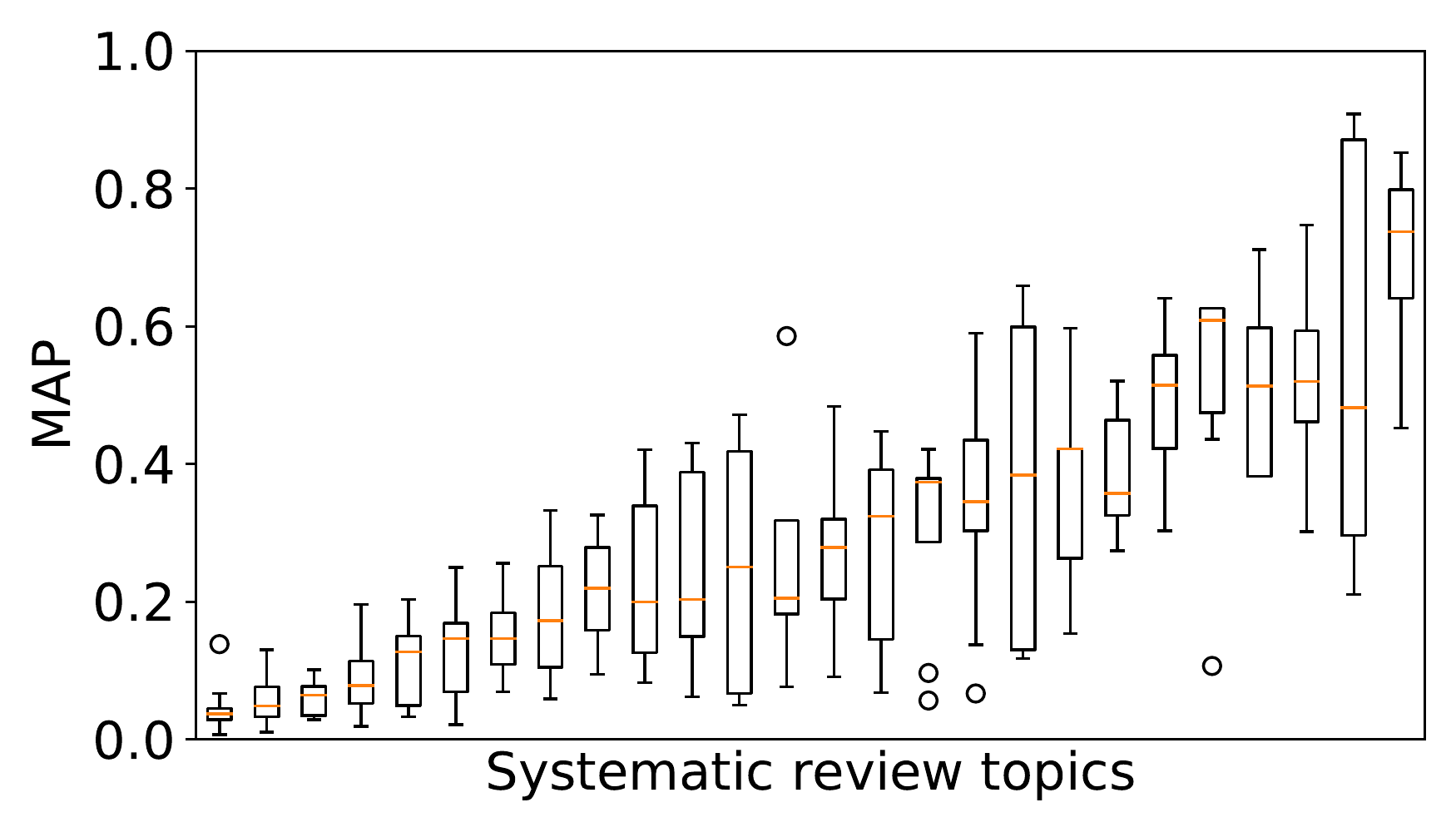}
	\caption{Single-SDR; 2019}
	\label{fig:results.2019var-single}
\end{subfigure}

\begin{subfigure}[t!]{.329\columnwidth}
	\includegraphics[width=\columnwidth]{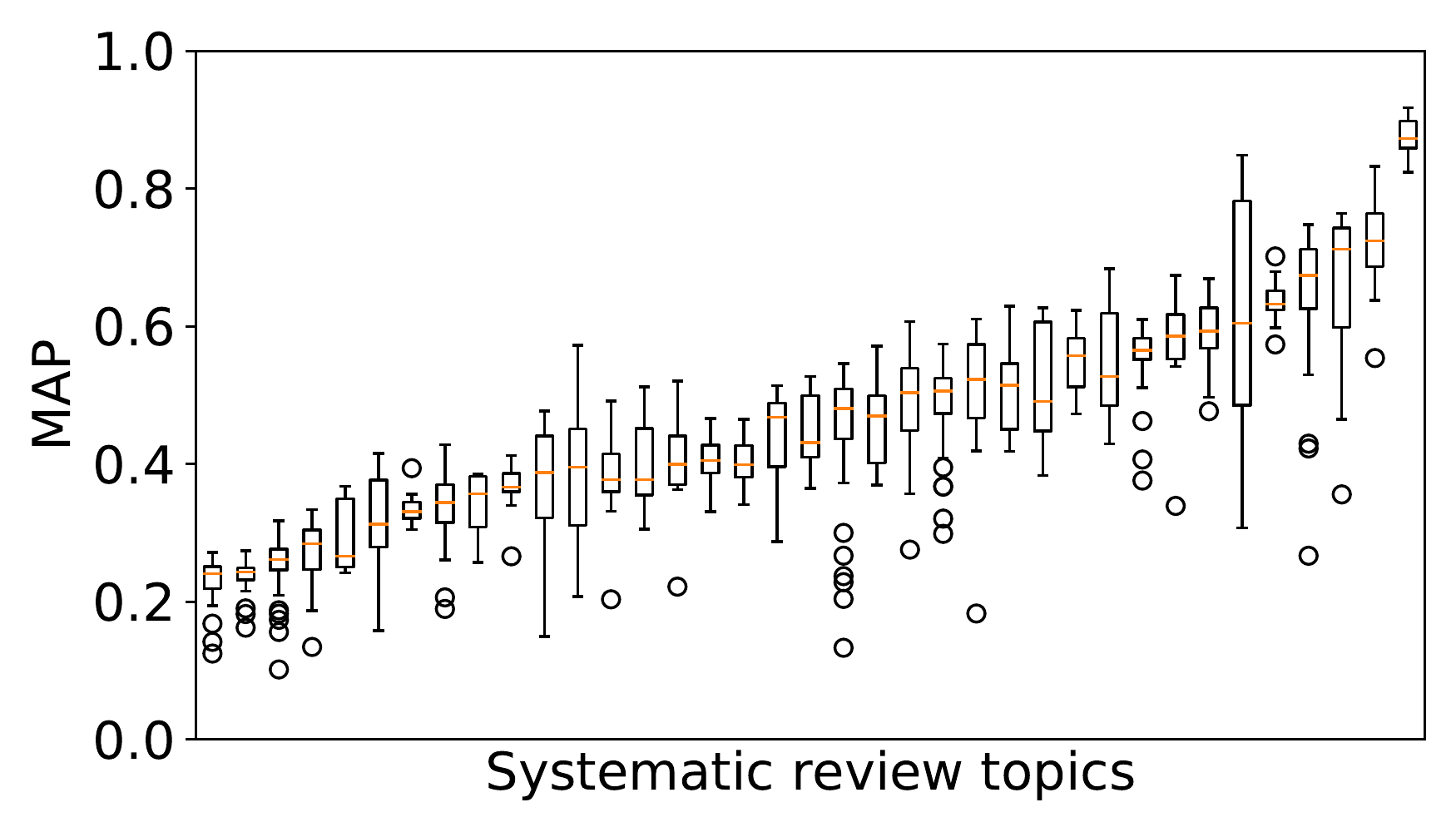}
	\caption{Multi-SDR; 2017}
	\label{fig:results.2017var-multi}
\end{subfigure}
\begin{subfigure}[t!]{.329\columnwidth}
	\includegraphics[width=\columnwidth]{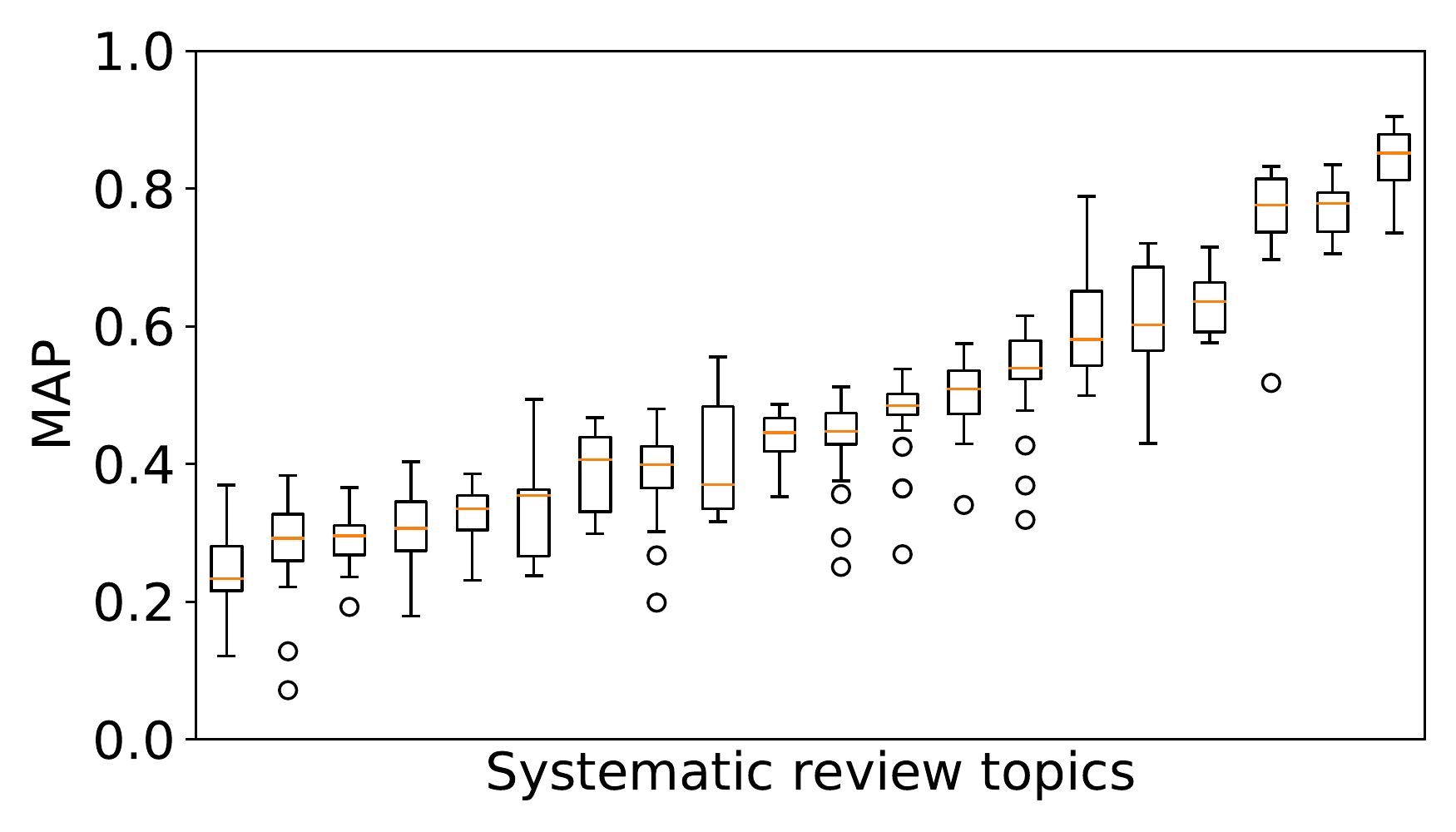}
	\caption{Multi-SDR; 2018}
	\label{fig:results.2018var-multi}
\end{subfigure}	
\begin{subfigure}[t!]{.329\columnwidth}
	\includegraphics[width=\columnwidth]{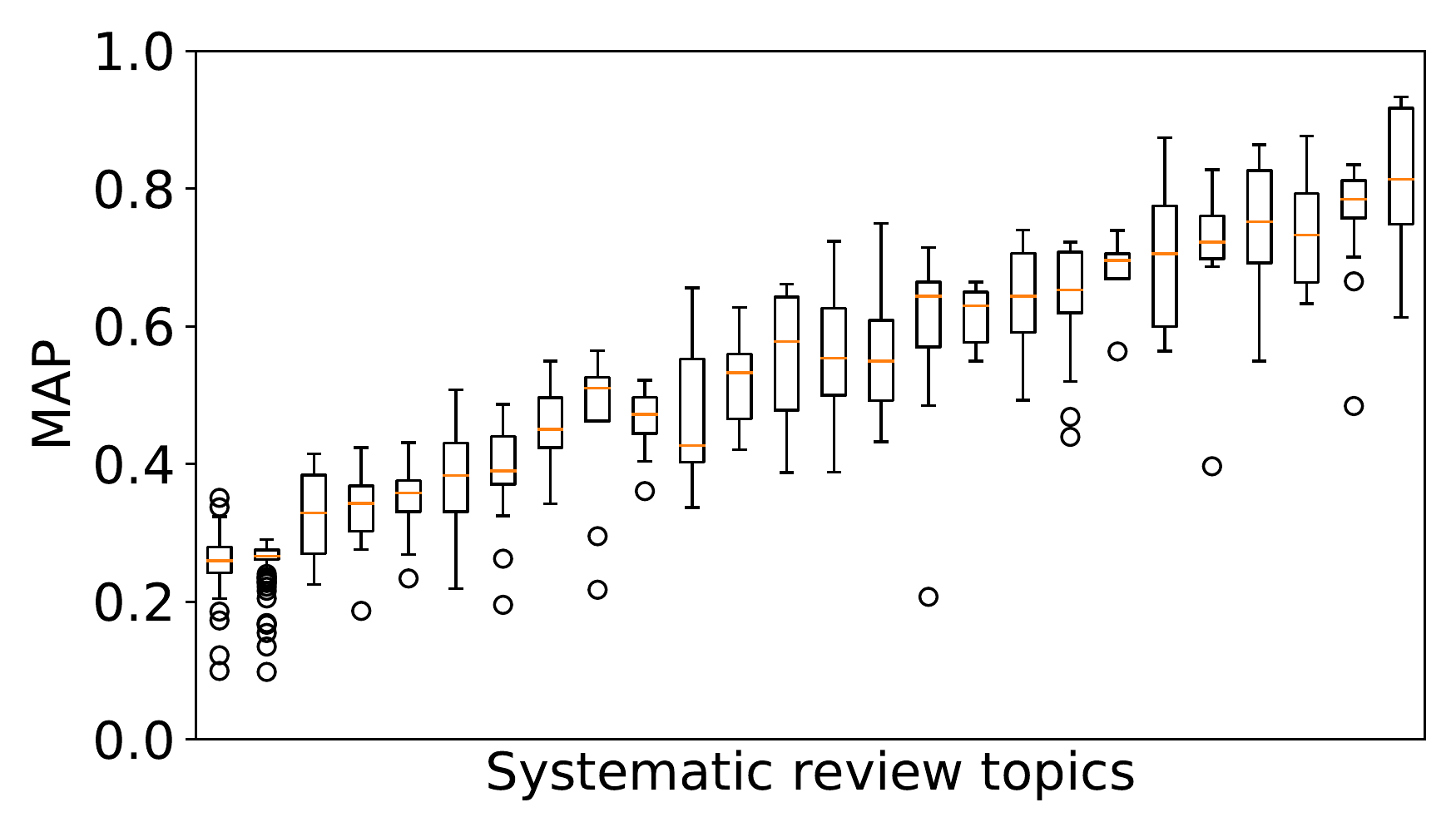}
	\caption{Multi-SDR; 2019}
	\label{fig:results.2019var-multi}
\end{subfigure}	
\caption{Topic-by-topic distribution of effectiveness (MAP) for the oracle-selected single-SDR-BOC-AES-P method (top figures) versus multi-SDR-BOC-AES-P.}
\label{fig:results.distribution}
\vspace{-12pt}
\end{figure}

Finally, we investigate the last research question: \textit{To what extent do seed studies impact the ranking stability of single- and multi-SDR?} We investigate this research question by comparing the topic-by-topic distribution of performance for the same results present in Table~\ref{table:results.multi}. These results are visualised in Figure~\ref{fig:results.distribution}. That is, we compare the multi-SDR results to the oracle single-SDR results, described in Section~\ref{sec:experimental-setup.comparing} so that we can fairly compare the variance of one to the other. 
We find that the variance obtained by multi-SDR is generally higher than that of single-SDR using DTA systematic review topics (Figure~\ref{fig:results.2017var-single} vs. Figure~\ref{fig:results.2017var-multi} --  and Figure~\ref{fig:results.2018var-single} vs. Figure~\ref{fig:results.2018var-multi}). We compute the mean variance across all topics, and find that the variance of multi-SDR (4.49e-2) is 10.89\% higher than single-SDR (4.44e-2) result for the 2017 dataset, and 11.76 \% for the 2018 dataset (single: 3.43e-2; multi: 4.17e-2).
For the 2019 dataset, we find that the variance of multi-SDR (7.93e-2) is 6.51\% lower than single-SDR (8.48e-2). 

However, when we randomly sample seed studies from each group for single-SDR, we find that the variance of multi-SDR is significantly lower: 53.2\% average decrease across 2017, 2018, and 2019. For space reasons, we do not include the full results. This suggests that the choice of seed study is considerably more important for single-SDR than for multi-SDR and that multi-SDR produces much more stable rankings, regardless of the seed studies chosen for re-ranking.

% oracle-single-sdr vs. multi-sdr variance
%2017: single: 0.004495233005946344 multi: 0.004991173695764913
%2018: single 0.0034358, multi: 0.0041756
%2019: single 0.008482662541326071 multi:  0.00793089683758504

% random-single-sdr vs. multi-sdr variance
%2017: single: 0.009487873819699556 multiple: 0.004991173695764913 . p_value: 0.000626014478743068
%13:42
%2018: single: 0.008880605976013346 multiple: 0.003967681777933726. p_value: 0.03168430616262175
%13:42
%2019: single: 0.018369295555980124 multiple: 0.00793089683758504. p_value: 0.003467571752057785
\vspace{-12pt}
\section{Related Work}
\vspace{-8pt}

Currently, it is a requirement for most high-quality systematic reviews to retrieve literature using a Boolean query~\cite{chandler2019cochrane,suhail2013methods}. Given that a Boolean query retrieves studies in an unordered set, it is also a requirement that all of the studies must be screened (assessed) for inclusion in the systematic review~\cite{chandler2019cochrane}. It is currently becoming more common for a ranking to be induced over this set of studies in order to begin downstream processes of the systematic review earlier~\cite{norman2019measuring}, e.g., acquiring the full-text of studies or results extraction. This ranking of studies has come to be known as `screening prioritisation', as popularised by the CLEF TAR tasks which aimed to automate these early stages of the systematic review creation pipeline~\cite{kanoulas2017clef,kanoulas2018clef,kanoulas2019clef}. As a result, in recent years there has been an uptake in Information Retrieval approaches to enable screening prioritisation~\cite{miwa2014reducing,chen2017ecnu,alharbi2017ranking,wu2018ecnu,alharbi2018retrieving,scells2017ltr,lee2018seed,lagopoulos2018learning,abualsaud2018system,zou2018technology,scells2020clf}. The vast majority of screening prioritisation use a different representation than the original Boolean query for ranking. Often, a separate query must be used to perform ranking, which may not represent the same information need as the Boolean query. Instead, the SDR method by Lee and Sun~\cite{lee2018seed} forgoes the query all together and uses studies that have a high likelihood of relevance, seed studies~\cite{suhail2013methods}, to rank the remaining studies. 
%These studies that have a high likelihood of relevance are called `seed studies', as they \textit{seed} the initial Boolean query formulation process~\cite{suhail2013methods}. 
These are studies that are known a priori to the query formulation step. The use of documents for ranking is similar to the task of query-by-document~\cite{yang2009query,lv2011learning} which has also been used extensively in domain-specific applications~\cite{kim2014diversifying,kim2014automatic,kim2015improving}. However, as Lee and Sun note, the majority of these methods try to extract key phrases or concepts from these documents to use for searching. SDR differentiates itself from these as the intuition is that the entire document is a relevance signal, rather than certain meaningful sections. Given the relatively short length of documents here (i.e., abstracts of studies), this intuition is more meaningful than other settings where the length of a document may be much longer.
\vspace{-12pt}
\section{Conclusions}
\vspace{-8pt}

We reproduced the SDR for systematic reviews method by Lee and Sun~\cite{lee2018seed} on all the available CLEF TAR datasets~\cite{kanoulas2017clef,kanoulas2018clef,kanoulas2019clef}.
Across all three of these datasets, we found that the 2017 and 2018 datasets share a similar trend in results than to the 2019 dataset. We believe that this is due to topical differences between the datasets and that proper tuning of SDR would result in results that better align with those seen in 2017 and 2018.
We also performed several pre-processing steps that revealed that the BOW representation of relevant studies could also share a relatively high commonality of terms compared to the BOC representation. Furthermore, we found that the BOC representation for SDR is generally beneficial and that term weighting generally improves the effectiveness of SDR.
We also found that multi-SDR was able to outperform single-SDR consistently. Our results also used an oracle to select the most effective seed studies to compare multi-SDR to single-SDR. This means that the actual gap in effectiveness between single-SDR and multi-SDR may be considerably larger.
Finally, in terms of the impact of seed studies on ranking stability, we found that although multi-SDR was able to achieve higher performance than single-SDR, multi-SDR generally had a higher variance in effectiveness.

For future work, we believe that deep learning approaches such as BERT and other transformer-based architectures will provide richer document representations that may better discriminate relevant from non-relevant studies. Finally, we believe that the technique used to sample seed studies in the original paper and this reproduction paper may overestimate the actual effectiveness. This is because a seed study is not necessarily a relevant study, and that seed studies may be discarded after the query has been formulated. For this, we suggest that a new collection is required that includes the seed studies that were originally used to formulate the Boolean query, in addition to the studies included in the analysis portion of the systematic review.

%Automating systematic reviews through screening prioritisation reduces the time it takes to create systematic reviews, resulting in improved health care delivery. 
Further investigation into SDR will continue to accelerate systematic review creation, thus increasing and improving evidence-based medicine as a whole.
%
% ---- Bibliography ----
%
% BibTeX users should specify bibliography style 'splncs04'.
% References will then be sorted and formatted in the correct style.
%
\bibliographystyle{splncs04}
\bibliography{bibliography}
\end{document}